\documentclass[10pt]{llncs}

\usepackage[all]{xy}
\usepackage{amsfonts}
\usepackage{amscd}
\usepackage{amsmath}
\usepackage{subfigure}
\usepackage{graphicx}
\usepackage{times}
\usepackage{url}
\usepackage{boxedminipage}
\usepackage{ifthen}

\newenvironment{keywords}{
       \list{}{\advance\topsep by0.35cm\relax\small
       \leftmargin=0cm
       \labelwidth=0.35cm
       \listparindent=0.35cm
       \itemindent\listparindent
       \rightmargin\leftmargin}\item[\hskip\labelsep
                                     \bfseries Keywords:]}
     {\endlist}

\newboolean{eps}
\setboolean{eps}{false}

\begin{document}


\title{We Can Remember It for You Wholesale: Implications of Data Remanence on the Use of RAM for True Random Number Generation on RFID Tags}

\author{Nitesh Saxena and Jonathan Voris}
\institute{Department of Computer Science and Engineering\\
Polytechnic Institute of New York University\\ 
Six MetroTech Center, Brooklyn, NY 11201\\
{\small \texttt{nsaxena@poly.edu}, \texttt{jvoris@isis.poly.edu}}
}

\maketitle
\begin{abstract}
Random number generation is a fundamental security primitive for RFID devices.
However, even this relatively simple requirement is beyond the capacity of today's
average RFID tag.  A recently proposed solution, Fingerprint Extraction and
Random Number Generation in SRAM (FERNS) \cite{FERNS,FERNS_journal}, involves the use of
onboard RAM as the source of ``true'' randomness. Unfortunately, practical
considerations prevent this approach from reaching its full potential. First,
this method must compete with other system functionalities for use of memory.
Thus, the amount of uninitialized RAM available for utilization as a randomness
generator may be severely restricted. Second, RAM is subject to \textit{data
remanence}; there is a time period after losing power during which stored data
remains intact in memory. This means that after a portion of memory has
been used for entropy collection once it will require a relatively extended
period of time without power before it can be reused. In a usable RFID based
security application, which requires multiple or long random numbers, this may lead to
unacceptably high delays.

\vspace{3.5mm}

In this paper, we show that data remanence negatively affects RAM based random number generation. We demonstrate the practical considerations that must be taken
into account when using RAM as an entropy source. We also discuss the
implementation of a true random number generator on Intel's WISP RFID tag,
which is the first such implementation to the authors' best knowledge. By
relating this to the requirements of some popular RFID authentication
protocols, we assess the (im)practicality of utilizing memory based randomness
techniques on resource constrained devices.


\end{abstract}

\section{Introduction}
\label{sec:intro}

The importance of Radio Frequency Identification (RFID) technology continues to
grow as RFID tags see deployment in an ever expanding variety of applications
and settings. Consequently, RFID security and privacy continues to be carefully
scrutinized by the research community. Providing security and privacy services
in RFID systems presents unique challenges due to the highly constrained nature
of RFID enabled devices. There has been much work on the development of
security and privacy mechanisms and protocols that attempt to take the limited
capabilities of RFID tags into account. Most, if not all, of these schemes
rely on the presence of one of the most fundamental cryptographic
primitives, random number generation.

While modern general purpose computers have many techniques available for the
generation of high quality random numbers, even this relatively simple
requirement is beyond the capacity of today's average RFID tag. The EPC air
interface specification for the most recent (Class 1 Generation 2) variety of
RFID tags includes a provision for pseudorandom number generation
\cite{epc_c1_g2}. The resulting random values are intended to be used only as a
collision prevention measure, however. When combined with the economic
considerations of these ultra-low cost devices, the values produced by these
generators are unlikely to be of high enough quality to be used as a source of
cryptographic randomness. 

A recently proposed alternative, Fingerprint Extraction and Random Number
Generation in SRAM (FERNS) \cite{FERNS,FERNS_journal}, involves the use of onboard RAM as the
source of ``true'' randomness. FERNS works by repurposing blocks of RAM into
physical fingerprints which, when run through a random number extractor (e.g.,
a hash function), produce random output. This technique is quite promising as
any device, regardless of its constraints, will contain some amount of onboard
memory from which randomness can be drawn. In addition to random number
generation, FERNS was also shown to be capable of creating unique fingerprints with which RFID tags can be uniquely identified. 

Unfortunately, practical considerations prevent the FERNS approach to random
number generation from reaching its full theoretical potential.
Since FERNS relies on preexisting memory circuitry as a source of entropy, it
must compete with other system functionalities for use of this shared resource.
Other code running on a RFID tag, such as the EPC protocol stack itself (that
is, the implementation of the protocol in software), will likely be occupying
the device's memory at any given point during execution. As such, the amount of
uninitialized RAM available for utilization as a randomness generator may be
severely restricted. Furthermore, RAM is subject to a phenomenon known as
\textit{data remanence}. While it is still volatile in the traditional sense,
due to properties of the underlying hardware such memory retains its contents
while receiving power and for a duration of several seconds afterwards. Thus,
there is a time period after losing power during which stored data remains
intact in memory. This means that after a portion of memory has been used
for entropy collection once, it will require a relatively extended period of
time without power before it can again be used in this capacity. In a usable
RFID based security application which requires multiple random numbers this
may lead to unacceptably high delays.

\paragraph{Our Contributions:} In this paper, we demonstrate the practical considerations that must be taken
into account when using RAM as an entropy source. We discuss the implementation of a true random number generator on
Intel's WISP RFID tag \cite{wisp1,wisp2}, which is the first such
implementation to the authors' best knowledge \cite{holcomb}. Using this as a
basis, we demonstrate how many bits of randomness one can expect to derive from
a RFID device's memory at a given time. Our results indicate that at most 309 bits of randomness can be derived from a tag with 512 bytes of RAM, with this figure dropping sharply as tag memory capacity decreases.

We then analyze the implications of
data remanence on RFID tags and the rate at which random number generation can
be performed. By relating this process to the requirements of some popular RFID
authentication protocols, we assess the (im)practicality of utilizing
memory based randomness techniques on resource constrained devices. As an example, we also
discuss the implications that RAM based randomness derivation would have on the
usage model of a typical RFID enabled access card. In addition,
we introduce potential attacks that could be launched on RFID system while
this method is in use.

\paragraph{Paper Organization:} The rest of this work is organized in the
following fashion. Section \ref{sec:background} introduces the fundamentals of
RFID systems and discusses related work. In Section \ref{sec:experiments}, our
experiments are explained in detail.  Section \ref{sec:discussion}
provides a discussion of the practicality of the studied approach, based on
our experiments. Finally, Section \ref{sec:conclusion} summarizes our results.

\section{Background}
\label{sec:background}

\subsection{RFID Overview}

RFID is an increasingly popular technology for computerized identification. An
RFID infrastructure consists of tags and readers. Tags are small transponders
that store data about their corresponding subject, such as an ID value.
Readers are used to query and identify these tags over a wireless channel. In most cases,
tags are passive or semi-passive, meaning they derive the power to transmit
data to a reader from the electromagnetic field generated when a reader sends a
query to a tag. Additionally, tags typically have memory only in the range of
32-128 bytes, perhaps just enough to store a unique identifier \cite{HB+}.
These ultra-low memory, computational, and power constraints are necessitated by
the fact that RFID tags are designed to be placed ubiquitously in consumer
products, appliances, and even users themselves (in the case of implanted payment tokens, for example). The minimalist
capabilities of these tags present unique privacy and security challenges, the issue of random
number generation being foremost among them. How can a device with limited
power, memory, computational capabilities, and user interfaces generate high quality
random numbers?

\subsection{WISP Tags}

In order to investigate this question, we utilized a special type of RFID tag
designed by Intel Research known as a Wireless Identification and Sensing
Platform (WISP) \cite{wisp1,wisp2}. WISPs are passively-powered RFID tags
that are compliant with the Electronic Product Code (EPC) protocol. Specifically, we
utilized the 4.1 version of the WISP hardware, which partially implements Class 1
Generation 2 of the EPC standard.  By following this standard and deriving
power only from the transmissions of a commercial RFID reader, WISPs closely
model the type of RFID tag one might expect to find in a typical contactless
access token.  Where the WISP differs from standard tags, however, is in its inclusion of an
onboard Texas Instruments MSP430F2132 microcontroller. This 16-bit MCU features
an 8 MHz clock rate, 8 kilobytes of flash memory, and 512 bytes of RAM. WISPs are the
first programmable passive RFID devices. Unlike standard RFID tags, which are
fixed function and state machine based, the flexibility of the WISP allowed us
to implement a random number generator and probe the behavior of memory on a
live, passive RFID device.

\subsection{Random Number Generation Based on RAM}
\label{sec:RNG}

A recent proposal to address the difficulty of generating random numbers on a
passive RFID device is called FERNS \cite{FERNS,FERNS_journal}. Instead of treating
uninitialized memory as a indeterminate blank slate, FERNS works by considering
this unused memory to be a fingerprint. This fingerprint can be used in two
complimentary ways. The first is as a means of identifying a given RFID tag
through the underlying physical characteristics of memory. The second is as a
potential source of entropy. The focus in this paper is on the latter application. Each unpowered RAM cell starts in an unstable
state, then moves to a stable state, representing either a `0' or a `1', once
supplied with power. Which of the two bit states the cell enters upon first
receiving power is dependant on the threshold voltage mismatch as well as the
thermal and shot noise of the cell. A large threshold voltage mismatch will
cause a RAM cell to reliably initialize to one bit value or the other. A small
mismatch, on the other hand, will be overshadowed by the cell's noise, causing
it to take on a value randomly at power up. It is the physical noise of these
RAM cells that supply entropy in the FERNS method.

Due to physical impurities, the random, well threshold matched cells will be
randomly scattered throughout the RAM. As these bits do not occur in convenient
proximity to one another, an extractor is necessary to pull these desirable
bits from the RAM sequence. A hash function can be used in this capacity. The
PH universal hash function of \cite{hashFunc} is recommended due to its
suitability for implementation in resource-limited hardware \cite{FERNS,FERNS_journal}. This
function is a variant of the NH hash function that was designed to be efficient
in software in order to accelerate the UMAC message authentication algorithm
\cite{UMAC}. PH is the result of a retooling of the NH function in order to
remove the need to perform carry operations, which makes the function more
suitable for a hardware implementation in terms of speed, space, and power
consumption.  PH is defined in Equation \ref{PH}. Blocks of uninitialized
memory are provided to the hash function as both key ($k_i$) and message ($m_i$) inputs; the output of
the hash function can then be used as a random bitstream.

\begin{equation}
  PH_{k}(m) = \displaystyle\sum_{i=1}^{8}(m_{2i-1} + k_{2i-1})(m_{2i} + k_{2i})
  \label{PH}
\end{equation}

\subsection{Data Remanence}

Since computer memory is volatile, it is a common belief that data stored in RAM
is completely lost as soon as it ceases to be supplied with power. This is not
entirely accurate, however. While the contents of unpowered RAM will certainly
degrade over time, the decay process takes several seconds to begin and several
more to complete \cite{SRAM_remanence,coldboot}. This process is due to the
low-level electrical components that comprise a RAM chip. In SRAM, for
instance, data is stored by setting the state of a flip-flop. This state is
maintained as long as the flip-flop continues to receive power
\cite{remanence}. This circuit does not lose its state immediately upon loss of
power, however. The state will remain for a short interval of time. Thus, there
is a brief time period after losing power during which stored data remains
intact. If power is again supplied before the end of this window the decay
process is halted. While the speed of data loss
varies greatly between individual chips, the rate of RAM decay is largely a function
of temperature. At high temperatures the degradation process is accelerated,
while if cooled to a low enough temperature the decay process can effectively be halted
indefinitely. 

The phenomenon of data remanence has serious repercussions for computer
security. Many times cryptographic data, such as keys, are stored in RAM. If an
adversary can gain physical access to a RAM chip containing sensitive material,
remove it, and read it on a different device before it fully decays, he or she
can potentially recover a full image of the memory contents, including any
stored secrets. Even if the decay process has already started to set in,
statistical techniques can be used to recover lost bits \cite{coldboot}. While data remanence plays an
important part in the work presented in this paper, it does not involve the recovery of data from
memory. Instead, we explore the implications of data remanence on the frequency of RAM
initialization.

\subsection{RFID Authentication}

One of the most important RFID security challenges is tag authentication. RFID tags are designed to
respond promiscuously to any query from a compatible reader. This behavior
makes the forging and duplication of RFID tags a relatively straightforward
process. Since tags respond to any query, there is little preventing an
adversary with the proper equipment from obtaining a tag's data, then creating
a new tag containing the exact same value \cite{HB+}. In many cases, traditional cryptographic solutions can not
be used by RFID tags due to their low computational and
memory capabilities. Several new solutions have been developed to address these
problems; one of the best known is the HB+ protocol \cite{HB+}. HB+ is a
challenge-response scheme based on the HB human authentication protocol
\cite{HB} that is designed with the computational and memory constraints of
RFID tags in mind, requiring only bitwise logic operators for computation. The
only other requisite of HB+ is the tag's capability to
generate high quality randomness, a property which today's RFID tags are ill
equipped to meet. The HB+ protocol requires at least
80 rounds \cite{HB_pound}, in each of which the RFID tag is expected to
generate a 224 bit random value, in order to attain an 80-bit security level. 
If these rounds are run in parallel \cite{katz}, a RFID tag will be required to produce 17,920 random bits at once.

Since its inception, various variants of HB+ have been proposed including HB++
\cite{HB++}, Trusted-HB \cite{Trusted-HB}, PUF-HB \cite{PUF-HB}, and HB\#
\cite{HB_pound}. Protocols derived from HB are not the only RFID security
mechanisms that require randomness to operate, though. Other protocols that are
based on pseudorandom functions will also require cryptographic random numbers
to be generated. For example, the tree based privacy-preserving authentication protocols of \cite{Molnar_library}
use pseudorandom functions that require high quality
randomness at each level of the tree.

\section{Experiments}
\label{sec:experiments}
In this section, we present the experiments used to measure the amount of randomness that can be derived from uninitialized RAM, as well as the rate at which this process can be performed, based on the practical limitations of RFID tags and their usage model.

\subsection{Experimental Setup}

\begin{center}
\begin{figure}[h]   
\begin{center}
 \includegraphics[height=33mm]{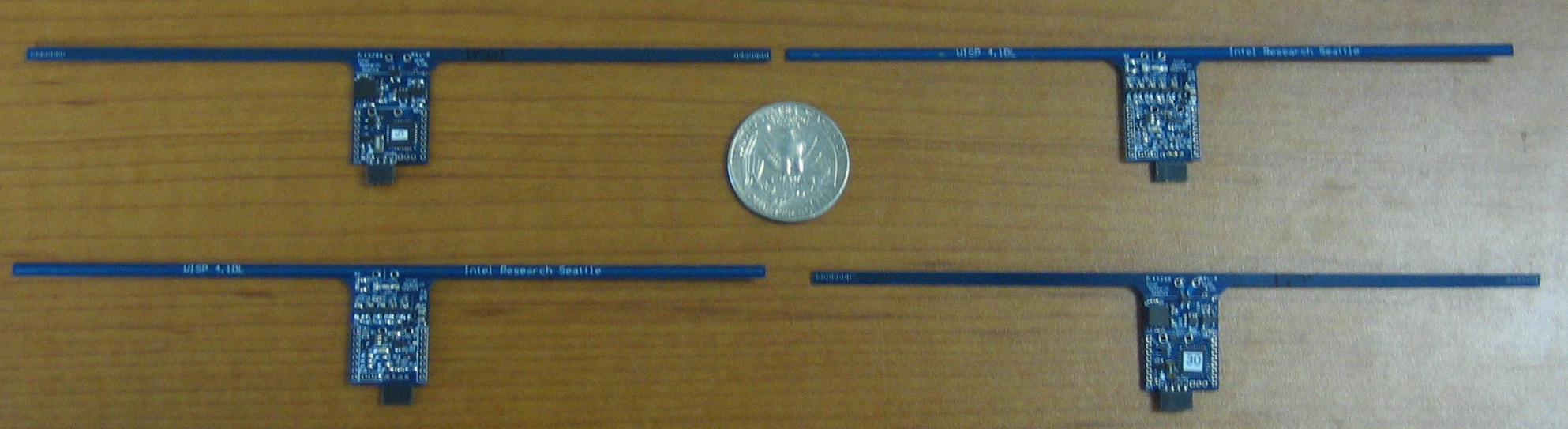}
\caption{{\footnotesize Four WISP 4.1 Tags with a U.S. Quarter Included for Scale \label{tag_pic}}}
\end{center}
\end{figure}
\end{center}

\begin{center}
\begin{figure}[h]   
\begin{center}
 \includegraphics[height=50mm]{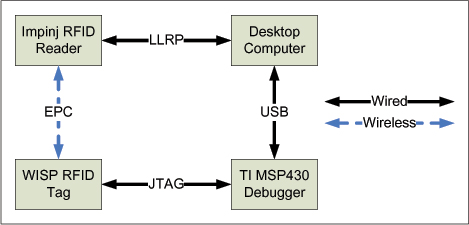}
\caption{{\footnotesize A Block Diagram of the Experimental Setup \label{block_diagram}}}
\end{center}
\end{figure}
\end{center}

We utilized the following configuration of equipment for our tests. Four WISP
tags of the latest hardware version, 4.1, were employed. The WISPs are depicted in Figure \ref{tag_pic}, with a U.S. quarter placed nearby to provide a sense of scale.
When these tags were
required to interact with the RFID reader they were loaded with the 6.0
revision of WISP firmware. At times when a tag's memory contents were of
interest, tags were  loaded with a C file containing nothing but a blank main function;
this was done in order to minimize the amount of RAM overwritten during program
execution. To program these WISP tags a Texas Instruments MSP-FET430UIF
debugging interface \cite{TI_debugger} was used, which was interacted with
through a desktop computer running the IAR Embedded Workbench IDE \cite{IAR}. The debugger was connected to the desktop machine with a USB cable and to the WISP tag over a JTAG interface.
We used an EPC compliant Impinj UHF Generation 2 Speedway RFID reader \cite{impinj}
running firmware version 3.2.1. Commands were issued to the reader from a
desktop machine through a custom application which communicated with the reader
over the Low Level Reader Protocol (LLRP). A block diagram of this hardware configuration is shown in Figure \ref{block_diagram}.


\subsection{WISP Implementation}

As a first step towards assessing the viability of deriving randomness from RAM
on a passive RFID device, we implemented the random number generation portion
of FERNS on a WISP tag. While RAM based randomness has been implemented on a
desktop computer, to our knowledge this is the first such implementation on a
passively-powered RFID device \cite{holcomb}. The implementation was done in C using the IAR
Embedded Workbench IDE. This was a relatively straightforward implementation of
the PH hash function (described in Section \ref{sec:RNG}). The main change that had to
be made was to alter the input block size of the PH hash function from 64 bits
to 16 bits, which reduced the size of the hash function's output from 133 bits
to 37 bits. This was done in order for the hash output to fit into a standard C
long long datatype. Had the original 133 bit output size been used, it would
have necessitated the use of a multiple precision arithmetic library, which would require the
dedication of the tag's scarce computational and memory resources.

Besides this practical consideration, reducing the block size of the PH
function has theoretical advantages as well. When this function's block size is
set to 64 bits, $64*32=2048$ bits of memory, half having been input as message
blocks and the other half having been treated as key blocks, are consumed to produce
a single 133 bit hash. When 16 bit blocks are used instead, only $16*32=512$
bits of RAM are needed to produce a 37 bit hash. While this decreases the
amount of random bits output from one call to the hash function, the smaller
block size allows PH to be called 3 more times on distinct blocks of
the same amount of uninitialized RAM, yielding  $37*4=148$ bits of randomness. Thus, reducing the
block size of the hash function allows more bits to be condensed from an
equivalent portion of memory. This would be problematic if the larger bit
amount exceeded the expected entropy of the values being hashed. Fortunately,
this is not the case. Each bit of tag memory is capable of yielding 0.103 bits
of entropy \cite{FERNS,FERNS_journal}. This means that the 148 bits output by multiple calls to PH with the
smaller hash function is still well within the 210 bits of entropy that 2048
bits of raw memory would be expected to produce.

As a preliminary test of the random values generated by this ``on-tag'' random number generator, 32 blocks of 16 bits each were read from an uninitialized area of the WISP's RAM. These values were hashed and written to a different memory address. We programmed the WISP to perform this operation once per query from the reader. The resultant 37 bit hash value was copied from memory into the WISP's EPC ID, which was then transmitted to the RFID reader in response to its queries. Surprisingly, we observed that identical values were being transmitted, indicating a clear lack of randomness. Since this random number generation technique is already known to work on traditional machines \cite{FERNS,FERNS_journal}, we set out to investigate the source of the discrepancy found on the WISP implementation. 

\subsection{Measuring Data Remanence on WISPs}

We altered the WISP tag's programming to transmit the contents of its memory to
the reader. This was accomplished by programming the tag to break its RAM into blocks. These blocks were then transmitted through the tag's EPC ID in the same manner as was done with the hash values. While there were occasional changes
in certain bytes, the contents of the memory seemed largely unchanged. This was
being caused by the WISP tag's retention of values between queries. Recall that
passive RFID tags derive power from reader queries. Thus, while continuously
being polled for hash or memory values, the WISP tag was receiving a continuous
supply of power, causing it to retain its RAM state rather than reinitializing
its memory after each query. 

We arranged a more thorough experiment to analyze the timing of data retention
on the WISP's memory. The methodology of our experiment was similar to that
employed in \cite{coldboot}. First, the WISP is connected to a desktop machine
using the MSP-FET430UIF debugging interface. We filled all
512 bytes of the WISP memory with a pseudorandom pattern generated on a desktop
machine using the Mersenne Twister \cite{MT} implementation included in the
random module of the Python programming language's standard library. This pattern was copied to the WISP's RAM
through the Embedded Workbench IDE. The WISP tag was then disconnected from the
debug interface, depriving it of power for a certain interval of time. After
this, the tag was reattached to the debugger. Rather than using the reader to
supply power to the tag and reading the memory values through the tag's EPC ID,
which is slow and prone to occasionally missing values, we resumed supplying
the tag with power over the debugger. The contents of the WISP's memory were
then read back. In order to calculate the tag's decay rate, we computed the
Hamming distance between the original pseudorandom pattern and the value read
back from the RAM. Two of the 512 bytes of RAM were always overwritten by the
debugger, so these bytes were left out of the analysis. The fact that the
original pattern was pseudorandom meant that it should contain an
approximately equal amount of each bit. Therefore, RAM was considered to be
fully decayed once the Hamming distance between the two strings was at or near
50\%. We did not alter the temperature of our tags; all tests were carried out at
room temperature.

\begin{center}
\begin{figure}[h]   
\begin{center}
 \fbox{ \includegraphics[width=115mm]{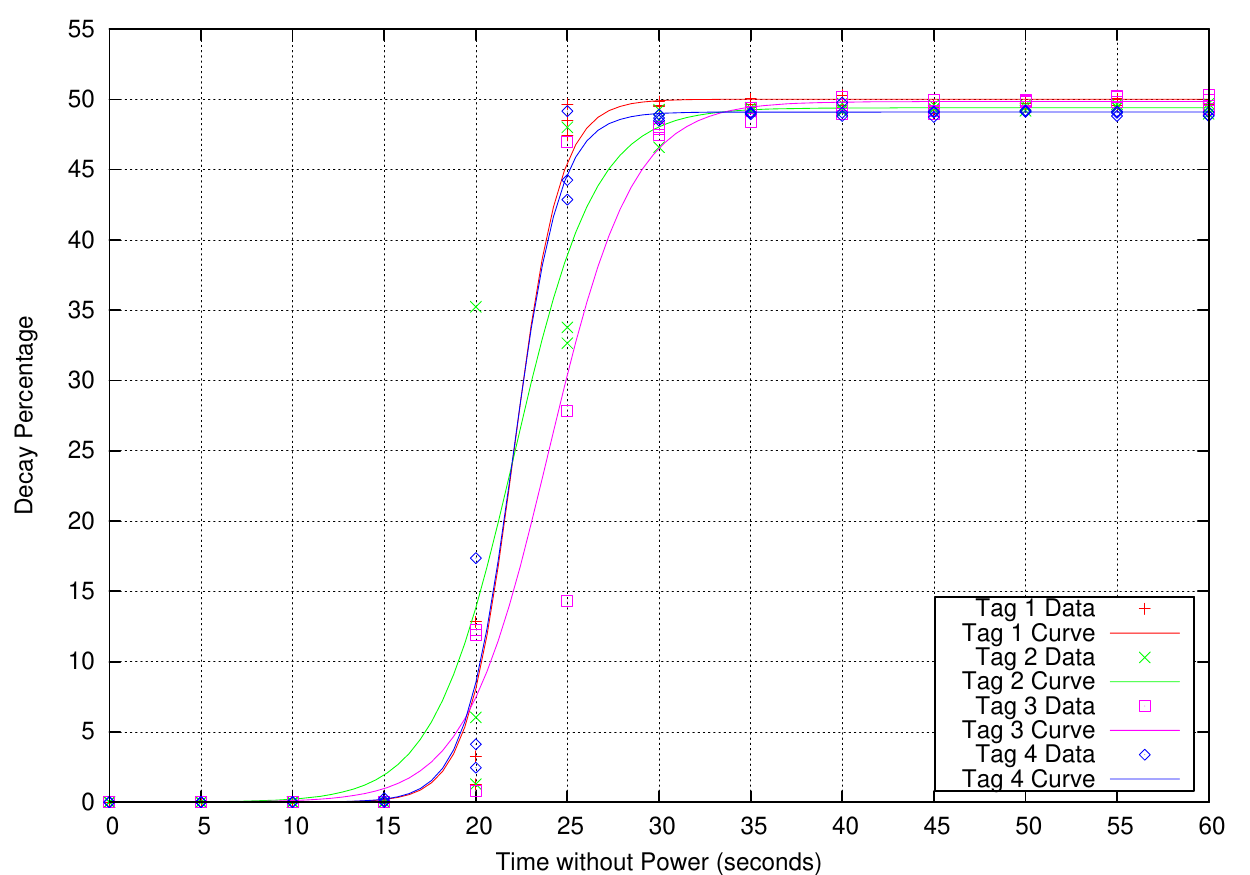}}
\caption{{\footnotesize Decay Rate for Each Tag \label{seperate_tags}}}
\end{center}
\end{figure}
\end{center}

\begin{center}
\begin{figure}[h]
\begin{center}
 \fbox{ \includegraphics[width=115mm]{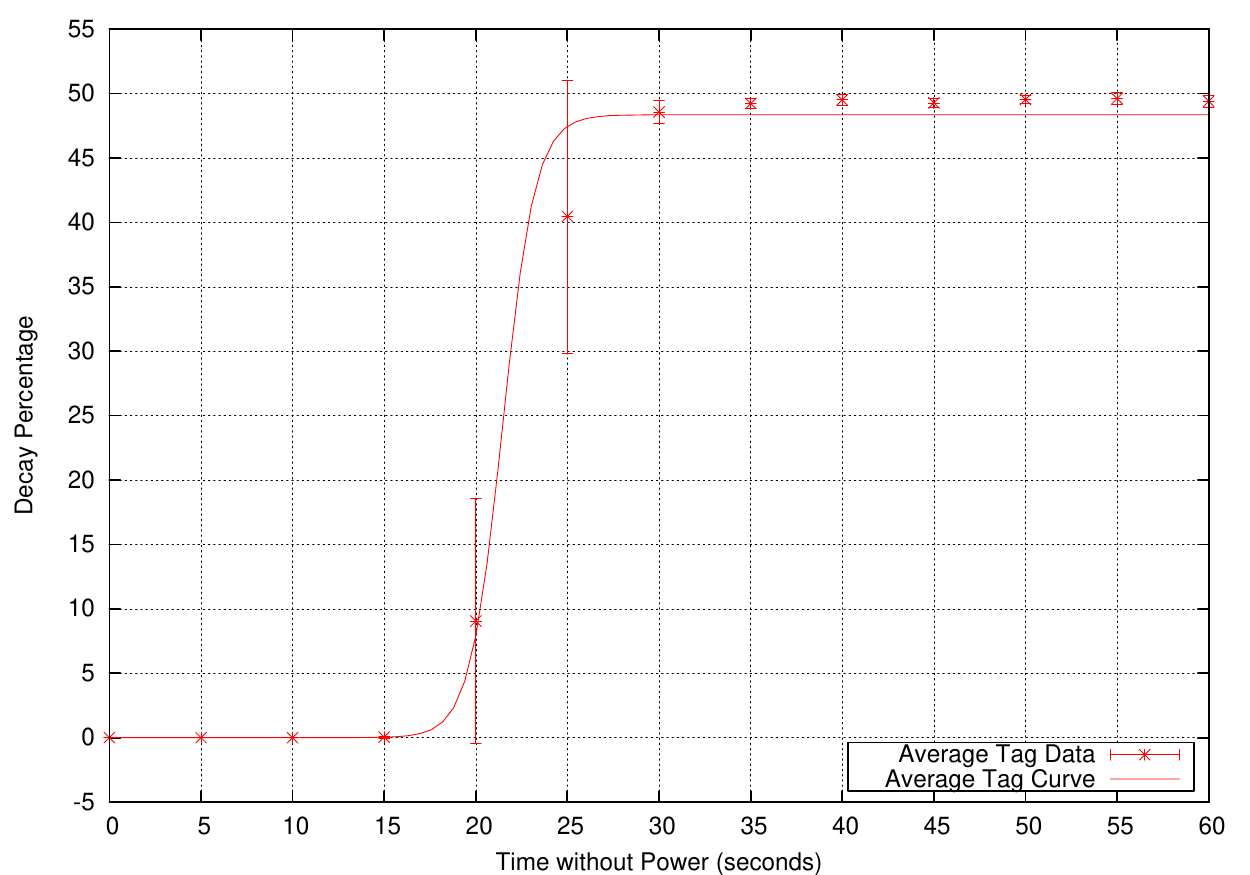}}
\caption{{\footnotesize Average Decay for All Tags \label{averages}}}
\end{center}
\end{figure}
\end{center}

We took samples after removing the WISP from power for a duration of 0 to 60
seconds at 5 second intervals. This test was performed on a population of 4
WISP tags. The results of our tests are shown in Figures \ref{seperate_tags}
and \ref{averages}. Figure \ref{seperate_tags} shows the per-tag decay rate,
while Figure \ref{averages} displays the average decay rate across all tags. A
logistic curve has been fit to each data set. While there were some minor
variations between samples, the decay rate observed on each tag was generally
well matched to this curve, showing an initial 15 seconds with little ($< 1\%$)
or no decay, then 15 seconds of very rapid decay, and concluding with the slow
decay of whatever data remained. From this data it is clear that, depending on
the particular tag, a powerless period of 25 to 30 seconds is required to allow the WISP's
512 bytes of memory to decay completely.

\subsection{Available Memory}

Having established how long it takes for a WISP's memory to return to an
uninitialized state, next we turned our attention to how much uninitialized memory
is available on a WISP at any given time. To determine the amount of unused RAM on the latest version (4.1) of WISP tags
used in our experiments, we loaded the tags with their default firmware and
then added the largest data structure the compiler would allow us to store in
the tag's RAM. We observed that the WISP protocol occupied 136 bytes of this
tag version's memory, leaving 376 bytes free for use as an entropy source.
Note, however, that by default this firmware does not implement all mandatory
aspects of the EPC standard. Enabling other features of the protocol in the
WISP firmware, such as the ability use multiple readers or read multiple tags
simultaneously, takes up an additional 12 bytes of RAM, leaving 364 bytes
available for random number generation. 

For earlier versions (2.0 and 2.1) of WISP tags, \cite{secret_handshakes} established that at any given time 112
bytes of WISP memory are occupied by the RFID protocol and stack. Assuming no
additional memory is used in order to program the tag with increased
functionality, this leaves a maximum of 144 uninitialized bytes for random
number generation. This assumes that no other RAM space is occupied by the authentication
protocol itself, which is unlikely to be true in practice.

\section{Discussion}
\label{sec:discussion}

\subsection{Practicality: Effects of Data Remanence and Available Memory}

Taking the HB+ and HB\# protocols as motivating examples, we ask: how feasible
is the use of RAM based random number generation for RFID applications in need
of random numbers? To provide 80 bit security, the HB+ protocol requires at
least 80 rounds \cite{HB_pound}, in each of which the RFID tag is expected to
generate a 224 bit random value. If these rounds are run in parallel, the WISP
must be capable of generating 17,920 random bits at a time. The randomness
requirements of the HB\# protocol are more modest, requiring a single round
where a 512 bit random value is generated by the tag, though this is at the cost of
a higher memory overhead. 

In the FERNS approach, as reported in \cite{FERNS,FERNS_journal}, an entropy rate of 0.103 bits of entropy per bit of
uninitialized memory was observed. Combining this with the maximum of 376 bytes
of unused RAM on a 4.1 WISP tag yields an expected random number capacity of 309
bits. A 4.1 WISP tag would therefore require its available memory to be hashed
58 times in order to meet the randomness requirements of the HB+ protocol and 2
times in order to generate enough randomness for the HB\# protocol. Since a
``cool down'' interval of about 30 seconds is required between memory hashes in order to allow a WISP
tag's RAM to return to its uninitialized state, this implies that 30 seconds of wait
time would be required for this type of tag to generate enough randomness for a single
HB\# session and 28.5 minutes of wait time would be necessary to create enough
random bits for one HB+ protocol instance. 

Of course, these estimates only apply to the latest iteration of WISP hardware.
RFID tags with lower capabilities would require even more time. On the earlier
2.0 or 2.1 versions of WISP tags, which featured 256 bytes of RAM in total, out
of which 144 bytes are available for hashing, 118 random bits could be expected
to be generated from each memory hash. This would necessitate 152 hashes for
HB+ and 5 hashes for HB\#, yielding uninitialization wait times of 76 minutes
and 2.5 minutes for each respective protocol. These figures are
specific to the specialized WISP hardware, which for the purposes of allowing
programming flexibility have memory capacities well beyond those of commercial
RFID tags.  A typical 5 to 10 cent RFID tag is expected to have a maximum of only 128 bits
of RAM in total \cite{HB+}, making the prospect of deriving sufficient
randomness from this source even dimmer.

\subsection{Effect on Usage Model}

The issue of RAM data retention is complicated by the RFID usage model.
For example, consider the case of contactless access card usage. Since cost
efficient tags are passively-powered, they power up when they come into range
of a compatible RFID reader and do not power down until they leave the reader's
field of view. This would mean that a standard RFID enabled access card would
have to be taken outside of the range of a reader in order to allow its memory
to ``cool down'' and return to an uninitialized state in order to perform random number
generation. Thus requiring multiple consecutive RAM hashes would significantly
alter the RFID usage model. Instead of a user presenting his or her tag to a
reader once, leaving it present momentarily, and returning the tag to a pocket
or other storage, users would have to repeatedly bring the access card within
the range of the reader and back out again, introducing a high user burden into
the authentication process. Further complicating the situation is the need for
the user to determine when to remove the card from reader's range and for
how long.  We suspect that specialized hardware could be added to a RFID tag
to address this problem by cutting power to memory after a random number
generation was requested. This would add complexity and thus cost to the tag,
however, contrary to their intended economic efficiency.  Furthermore, a
hardware based solution would also not address the underlying need to wait for
several seconds between two consecutive RAM reads.

\subsection{Potential Attacks}

The need to move a tag outside of the range of a reader for a fresh random
number generation also introduces the potential for new attacks. If an
adversary were able to continuously supply power to a tag which made use of
its RAM for randomness purposes, he or she would essentially force the tag to
continuously reuse the same RAM values for hashing. This would make the values
generated extremely predictable, undermining the security of any authentication
scheme or cryptographic protocol built on top of the random number generator.  As mentioned above, hardware
could be added to lock down a tag's memory until it has time to return to a
decayed state. However, this would create the potential for a denial of service
(DoS) attack where an attacker continuously powered a tag, preventing it from
generating any random numbers and thus from being used at all. While DoS attacks on RFID
systems are always possible by simply jamming the radio signals involved, this type of
attack is worse in the sense that it does not involve any jamming in the traditional
sense. All an attacker would need to do is repeatedly issue queries
to the tag, rather than continuously jam an entire portion of the radio
spectrum.

\section{Conclusion}
\label{sec:conclusion}

To conclude, we have presented several practical shortcomings of using general
purpose memory as a source of randomness for low cost RFID devices. Since RAM
is already in short supply on such resource constrained devices, much of it
will likely be in use and thus unavailable as a source of randomness. Due to
the phenomenon of data remanence, a longer than expected wait time is required
between consecutive uses of RAM as an entropy source, making its repeated
utilization impractical in the RFID usage model. 

We do not conclude, however, that RAM based randomness derivation should be
discarded. This innovative technique remains attractive due to its repurposing
of existing hardware, which is important for minimizing the costs of tag
production. On its own, however, this method seems unlikely to be able to
handle the randomness requirements of current RFID authentication protocols
such as HB+, HB\#, and related variants. 

In practice, many services derive random numbers from environmental noise. As
future work, we plan on investigating the viability of alternative sources of
randomness, such as onboard sensors, to collect ambient noise of different
forms. This approach would not be subject to the time and space constraints faced when
harvesting entropy from memory. As sensing platforms, WISP tags are well suited
to exploring this area. For example, the current 4.1 iteration of WISP
hardware features an onboard accelerometer, temperature sensor, voltage sensor,
and capacitance sensor. Additionally, it is possible to add new sensors by
wiring them to a WISP. We intend to analyze ways in which entropy sources such
as these can be aggregated to efficiently produce the amount of randomness necessary to support various cryptographic protocols aimed at low cost tags.

\subsubsection{Acknowledgements.}

The authors would like to thank Dan Holcomb, David Molnar, and the anonymous RFIDSec'09 reviewers for their helpful feedback on an earlier version of this paper. We would also like to give a special thanks to Dan Yeager both for his comments on this paper and his advice on WISP programming.

\bibliographystyle{abbrv}   
{
\bibliography{paper}


\end{document}